\documentclass[10pt, conference, letterpaper]{IEEEtran}
\usepackage{amsmath}
\usepackage{amssymb}
\usepackage{booktabs}
\usepackage{cite}
\usepackage{graphicx}
\usepackage[caption=false,font=footnotesize]{subfig}
\usepackage{multirow}
\usepackage{courier}
\usepackage{url}

\newcommand{\artifacturl}{{\fontfamily{pcr}\fontseries{m}\fontshape{n}\selectfont https://anonymous.4open.science/r/\allowbreak wirelessopsbench-artifact-D969/}}
\usepackage[percent]{overpic}
\usepackage[table]{xcolor}
\definecolor{opsaccent}{RGB}{132,207,196}
\definecolor{opsrowgray}{gray}{0.92}
\newcommand{\relatedworkhead}[1]{\par\smallskip\noindent\makebox[\columnwidth][c]{\textbf{#1}}\par\nobreak\smallskip}

\title{WirelessOpsAgent: A Benchmark and Agent Design for Action Assurance in Wireless Networks}

\author{
\IEEEauthorblockN{Zijian Lu\IEEEauthorrefmark{1}, Yiping Zuo\IEEEauthorrefmark{1}, Hao Xu\IEEEauthorrefmark{2},
Weicong Chen\IEEEauthorrefmark{2}, Xin He\IEEEauthorrefmark{1}, Jiajia Guo\IEEEauthorrefmark{2}, and Shi Jin\IEEEauthorrefmark{2}}
\IEEEauthorblockA{\IEEEauthorrefmark{1}Nanjing University of Posts and Telecommunications, Nanjing, China\\
18818732360@163.com, \{zuoyiping, xhe\}@njupt.edu.cn\\
\IEEEauthorrefmark{2}National Mobile Communications Research Laboratory, Southeast University, Nanjing, China\\
\{hao.xu, cwc, jiajiaguo, jinshi\}@seu.edu.cn}
}

\begin{document}

\maketitle

\begin{abstract}
Large language model (LLM) agents are emerging as planners for autonomous wireless network operations. Yet a task answer that is correct at proposal time can still be unsafe at execution time if supporting telemetry is stale or inconsistent. Existing benchmarks, however, mainly evaluate task solving from fixed observations and leave support checking at execution time untested. We introduce WirelessOptBench, a benchmark for action assurance in wireless operations. It turns wireless tasks into execution state decision episodes with controlled telemetry faults and action constraints. We further develop WirelessOpsAgent, which grounds candidate actions in current evidence and repairs recoverable support failures before execution. Across three backbone evaluations with 600 episodes each, WirelessOpsAgent achieves up to 0.983 Exact Action Accuracy. On Claude Sonnet 4.6, the Unsafe APPLY Rate decreases from 82.2\% to 10.3\% relative to the safest baseline. We make WirelessOptBench available at \artifacturl.
\end{abstract}


\section{Introduction}
Modern wireless networks expose increasingly programmable control loops. Network slicing and Open Radio Access Network (O-RAN) architectures allow software to reconfigure radio resources, service policies, and mobility actions, while intent based management lets operators specify objectives at a higher level~\cite{leivadeas2023intentBasedNetworking,foukas2017networkSlicing,garciaSaavedra2021oran}. These interfaces broaden automated control, but they also make operation depend on telemetry collected across distributed network functions and updated at different rates. A control decision must therefore account for both the service objective and the network state at execution. Large language model (LLM) agents are attractive in this setting because they can connect operator intent, wireless reasoning, and tool use within one workflow.

\begin{figure}[!t]
\centering
\includegraphics[width=\linewidth]{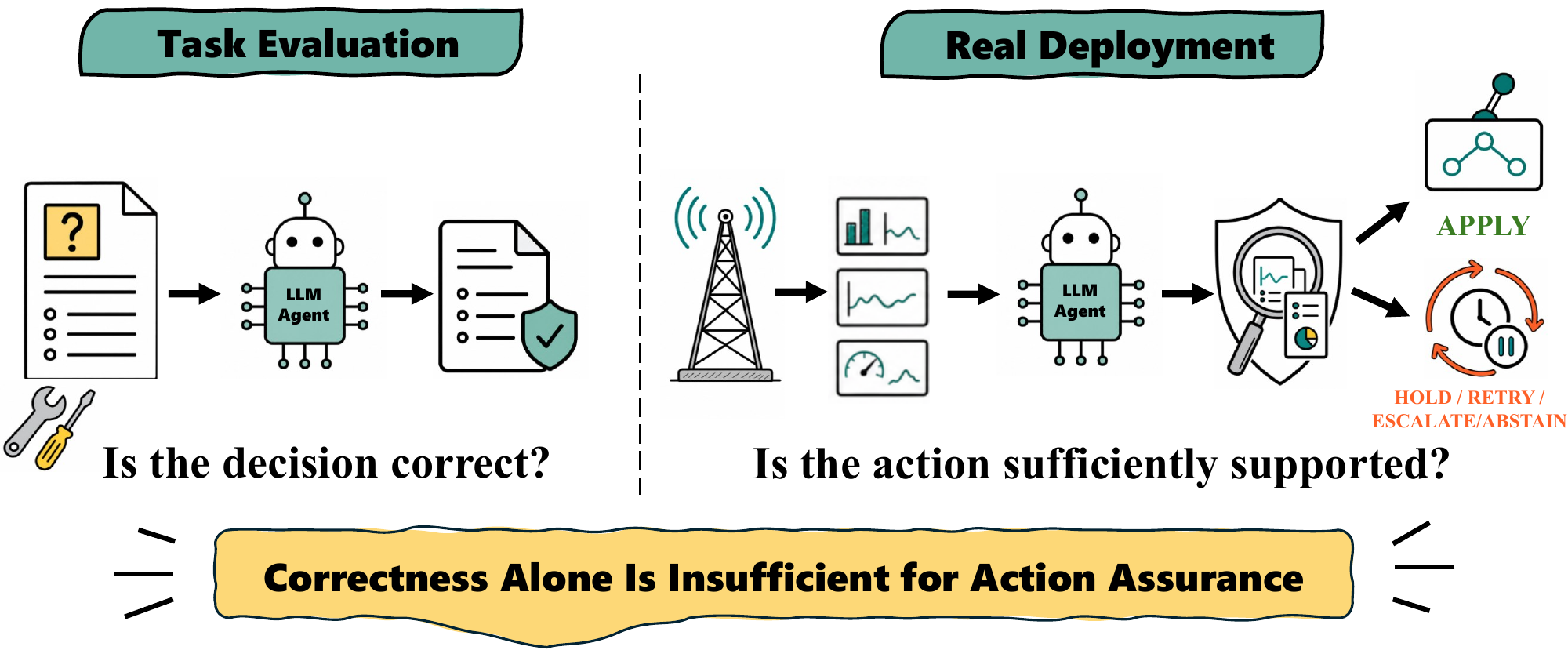}
\caption{Task evaluation checks decision correctness. Action assurance also checks whether current evidence supports execution.}
\label{fig:motivation}
\end{figure}

In these loops, support can expire between proposal and execution. Large scale monitoring studies show that coverage gaps can hide operational failures~\cite{ding2024rdprobe}, while mitigation ranking improves when it accounts for predicted end to end impact~\cite{namyar2025enhancing}. These findings align with classical results showing that distributed software-defined networking (SDN) correctness depends on state consistency~\cite{sakic2018adaptiveconsistency} and that even correct endpoint configurations can traverse incorrect intermediate states during updates~\cite{reitblatt2012abstractions}. Freshness, consistency, coverage, and availability therefore determine whether a previously justified decision remains admissible when execution begins.

Despite this concern, recent LLM agent benchmarks mainly test task solving. WirelessAgent connects wireless perception and planning to tool assisted action generation, and WirelessAgent++ further automates workflow construction~\cite{tong2026wirelessagentjournal,tong2026wirelessagentautomatedagenticworkflow}. WirelessBench evaluates reasoning, network slicing, and mobility service assurance~\cite{tong2026wirelessbenchtoleranceawarellmagent}. Configuration oriented agents also generate or repair network configurations with simulators and formal verification~\cite{wang2024netconfeval,protogeros2026benchmarking}. This body of work establishes whether an agent can derive an appropriate answer or plan from a well formed input. The supplied observations usually remain fixed task context, and assessment stops before the proposed action reaches the control plane.

At that boundary, task correctness is only the first gate. An allocation computed from current load, channel quality indicator (CQI), and bandwidth may be valid when proposed, yet unsafe after one measurement expires. The same proposal may need to wait for a newer observation, retry an unavailable source, or be referred to an operator when trusted sources disagree. Schema changes, unit mismatches, and out of order updates create a similar problem: the final answer can remain plausible even though one critical field has lost valid support. Fig.~\ref{fig:motivation} illustrates the separation between solving the wireless task and authorizing the resulting action.

These cases define the need for \textit{action assurance}: checking whether a proposed action still has valid support from the network state. Agent risk constraints, guarded control, and functional verification can prevent some unsafe actions~\cite{yuan2024rjudge,wang2025agentspec,seto1998simplex,mitsch2016modelplex}, but they do not show that every critical field is backed by admissible observations. Safety prompts and anomaly scores have the same limitation. Response revision methods such as Self-Refine and CRITIC can revisit a candidate through self feedback or tool interaction~\cite{madaan2023selfrefine,gou2024critic}. In network control, rewriting the whole response may disturb fields that are already valid, while refusing every irregular case discards repairable schema aliases, unit conversions, and verified recomputations.

WirelessOpsAgent implements this check in two stages. The first binds the public observations into a canonical ledger before generation. This improves the input but cannot certify the later answer or plan. The second treats the output as a proposal, links its fields to their support, and checks it before release. Recoverable failures are repaired only inside the affected dependency region, and the rebuilt state is revalidated. Each decision carries a record of the observations used, accepted repairs, and unresolved risk.

To evaluate this boundary, WirelessOptBench preserves the original wireless questions while varying the execution state. Each episode couples the question with a task specific evidence graph, an execution contract, a stress realization, and a rule based reference action. The nominal answer stays fixed. The realized state determines whether the action is executable. This separates task solving from authorization and avoids assuming that the final split contains every stress realization for each base item. WirelessOptBench measures this problem before execution, while WirelessOpsAgent addresses it in the agent loop. The main contributions are:
\begin{itemize}
\item We introduce WirelessOptBench, an action assurance benchmark that converts wireless reasoning, network slicing, and mobility service assurance tasks into execution state decision episodes with evidence graphs, execution contracts, telemetry stresses, and rule based reference actions.
\item We design WirelessOpsAgent, a closed loop wireless network agent that treats model outputs as proposals, grounds critical fields in network evidence, diagnoses integrity violations, repairs affected dependencies, and authorizes actions only after revalidation.
\item We provide an auditable evaluation protocol based on observable decision bundles and replayable authorization records, enabling matched measurement of action correctness, safe progress, and unsafe execution across backbones and baselines.
\end{itemize}

\section{Related Work}
\label{sec:related}

This work connects three lines: network agents, execution guards, and operations benchmarks. The common issue is whether supporting observations remain valid as agents move from answers to actions.

\relatedworkhead{Agents for Network Operations}
Tool using agents interleave reasoning with external observations~\cite{yao2023react,schick2023toolformer,qin2024toolllm}. WirelessAgent introduced LLM based wireless management, and WirelessAgent++ searches for workflows over calculation, slicing, and mobility tasks~\cite{tong2026wirelessagentjournal,tong2026wirelessagentautomatedagenticworkflow}. NetConfEval and Cornetto connect configuration generation or repair to network analysis~\cite{wang2024netconfeval,protogeros2026benchmarking}, while Self-Refine, CRITIC, and AgentSpec improve revision, critique, and constraint enforcement~\cite{madaan2023selfrefine,gou2024critic,wang2025agentspec}. However, longer tool chains depend on observations that may become stale, inconsistent, or detached from their derivations. Revision or behavioral enforcement alone does not restore current field level support.

\relatedworkhead{Programmable and Verifiable Network Operations}
Intent based management translates goals into policies, slicing exposes service specific resource control, and FlexRAN and O-RAN open programmable radio interfaces~\cite{leivadeas2023intentBasedNetworking,foukas2017networkSlicing,foukas2016flexran,garciaSaavedra2021oran}. Wireless controllers then adapt decisions from time varying measurements~\cite{sun2019mlWireless}. Header Space Analysis, NetPlumber, VeriFlow, and Batfish check network invariants~\cite{kazemian2012hsa,kazemian2013netplumber,khurshid2013veriflow,fogel2015batfish}. Simplex and ModelPlex guard controller execution~\cite{seto1998simplex,mitsch2016modelplex}. The World Wide Web Consortium (W3C) provenance (PROV) data model represents data lineage~\cite{moreau2013provdm}. These mechanisms generally assume observations have already been mapped into a reliable control state. Our setting instead checks that mapping when an action is released.

\relatedworkhead{Network Operations Datasets and Benchmarks}
Existing network operation benchmarks fall into two main settings. TeleLogs and AgenticOpsEval evaluate diagnosis from fixed logs, metrics, and traces, including fifth-generation (5G) root cause identification~\cite{sana2025reasoning,cai2026multidataset}. NetConfEval and Cornetto evaluate configuration generation or repair from static network inputs~\cite{wang2024netconfeval,protogeros2026benchmarking}. WirelessBench instead evaluates wireless calculation, slicing, and mobility answers from a given observation set~\cite{tong2026wirelessbenchtoleranceawarellmagent}. These settings are valuable, but they either treat evidence as an offline snapshot or focus on solving the wireless task itself. They do not test the execution boundary where the task and candidate decision stay fixed while the supporting observations change. WirelessOptBench targets this boundary with execution state observations and an explicit authorization contract.

\section{WirelessOptBench}
\label{sec:benchmark}

\begin{figure*}[!t]
\centering
\includegraphics[width=0.99\textwidth]{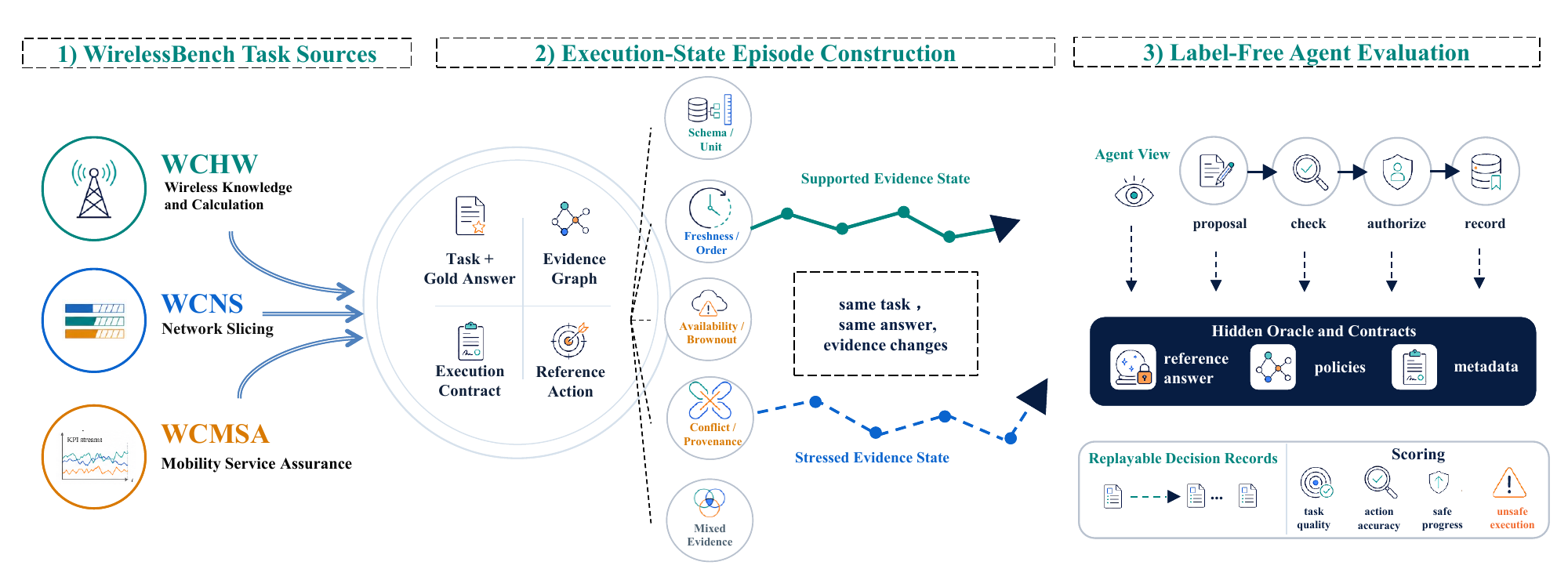}
\caption{WirelessOptBench pairs fixed wireless tasks with execution state observations and scores decisions against state derived reference actions.}
\label{fig:benchmark-construction}
\end{figure*}

The benchmark measures the transition from a plausible wireless solution to an executable operational decision. WirelessBench provides the underlying wireless tasks~\cite{tong2026wirelessbenchtoleranceawarellmagent}. TeleLogs contributes annotated 5G troubleshooting cases, while AIOps2025 and RCA100 provide expert labeled failures over multimodal service telemetry~\cite{sana2025reasoning,cai2026multidataset}. These public sources ground the operational fault vocabulary and anti shortcut design. Because they do not model whether a proposed action can be executed, we convert the inherited tasks into new execution state decision episodes that preserve task semantics while changing the available observations. Fig.~\ref{fig:benchmark-construction} summarizes this conversion.
\begin{table}[htbp]
	\centering
	\caption{WirelessOptBench stress set composition (episodes). Wireless knowledge and calculation (WCHW), 5G network slicing (WCNS), and mobility service assurance (WCMSA) denote the three task families.}
	\label{tab:bench-composition}
	\scriptsize
	\begin{tabular*}{\columnwidth}{@{\extracolsep{\fill}}lrrrrrr@{}}
		\toprule
		Task & Schema & Fresh. & Avail. & Conflict & Mixed & Total \\
		\midrule
		WCHW & 40 & 40 & 40 & 40 & 40 & 200 \\
		WCNS & 40 & 40 & 40 & 40 & 40 & 200 \\
		WCMSA & 40 & 40 & 40 & 40 & 40 & 200 \\
		\midrule
		Total & 120 & 120 & 120 & 120 & 120 & 600 \\
		\bottomrule
	\end{tabular*}
\end{table}
\subsection{Wireless Execution Model}
WirelessOptBench ties executable actions to radio checks at the task level. Consider user $u$ served by base station $b$, with $j$ indexing other base stations. Its signal-to-interference-plus-noise ratio (SINR) at execution time is
\begin{equation}
\gamma_u(t)=\frac{P_b h_{u,b}(t)}
{N_0+\sum_{j\ne b}P_j h_{u,j}(t)}.
\label{eq:sinr}
\end{equation}
Here, $P_b$ is transmit power, $h_{u,b}(t)$ is channel gain, and $N_0$ is noise power. With bandwidth $W_u(t)$, the achievable rate is
\begin{equation}
r_u(t)=W_u(t)\log_2\!\left(1+\gamma_u(t)\right).
\label{eq:rate}
\end{equation}
This gives the calculator check for wireless questions. For a slicing plan, the assigned bandwidth must fit the cell budget $W$, and each admitted user must meet its minimum rate:
\begin{equation}
\sum_u W_u(t)\leq W,\qquad r_u(t)\geq r_u^{\min}.
\label{eq:slicing}
\end{equation}
A mobility plan uses a separate handover condition. The target base station $j$ must sustain a reference signal received power (RSRP) margin $H$ for trigger time $T$:
\begin{equation}
\mathrm{RSRP}_{u,j}(\tau)-\mathrm{RSRP}_{u,b}(\tau)\geq H,
\quad \tau\in[t-T,t].
\label{eq:handover}
\end{equation}
These checks are unsafe when their measurements have expired. If field $k$ was observed at $t_k$ and has freshness limit $\delta_k$, its age must satisfy
\begin{equation}
\Delta_k(t)=t-t_k,\qquad \Delta_k(t)\leq\delta_k.
\label{eq:evidence-age}
\end{equation}
A derived value inherits the largest age among its inputs. Each episode instantiates only checks supported by its public fields, so an apparently feasible plan cannot bypass stale or unavailable radio state.

\subsection{Benchmark Construction}

With the execution checks fixed, WirelessOptBench samples 200 unique base items from each of three WirelessBench task families: wireless knowledge and calculation (WCHW), 5G network slicing (WCNS), and mobility service assurance (WCMSA). WCHW is restricted to items covered by tools for which formula lookup and calculator observations can be generated.

Each base item retains its question and reference answer, while an observation ledger, an execution contract, and a stress realization define the execution state episode. The ledger exposes metadata and dependencies. The contract specifies the fields and network constraints required before execution.

\subsection{Reference Answers and Action Labels}
\label{sec:benchmark-labels}

The evaluator must distinguish the nominal wireless answer from the execution decision. Each episode therefore retains the WirelessBench reference answer and adds a four action reference policy. The agent output space contains five actions because ABSTAIN remains available as a fallback, but it is never a reference action. A predicted ABSTAIN is retained as an incorrect action rather than dropped. For APPLY, WCHW requires only the reference answer, while WCNS and WCMSA also require a contract compliant allocation or mobility plan. The evaluator recomputes plan completeness, capacity or quality of service (QoS) feasibility, and action budget compliance.

Labels follow an ordered policy over six predicates. Ambiguous schema, material conflict, or independent failures yield ESCALATE. Otherwise, a required source in a retryable outage yields RETRY, unknown critical schema or stale required state yields HOLD, and no blocker yields APPLY. The policy reads the realized state and contract rather than the stress domain name, allowing both safe and blocking cases within a domain.

The generator receives no requested action. After realization, a fact level oracle and a public view certificate checker must agree, and every blocking label must cite blocking observations. Deterministic selection retains 600 of 7,680 candidates with no overlap between development and final anchors and a label distribution of 200 APPLY, 100 HOLD, 100 RETRY, and 200 ESCALATE. Their agreement is an internal consistency check because both paths implement the same benchmark specification.

\subsection{Fairness and Audit}
All methods receive the same frozen episodes: task input, public observation ledger, metadata, action semantics, and execution contract. Although the public contract defines the meanings of APPLY, HOLD, RETRY, and ESCALATE, the realized reference action is never exposed. WirelessOpsAgent and the baselines see no hidden stress identifiers, oracle certificates, injected originals, reference answers, or reference actions. Differences are limited to how each workflow inspects, revises, and authorizes its proposal, so the governor is evaluated on reconstructing authorization state from public evidence rather than on access to evaluator labels.

The suite combines 200 reference APPLY hard negatives with 400 supported blocking cases across 32 realization families, so neither always applying nor always blocking is rewarded. Across 1,200 label only mutations, changing the hidden action leaves the public view and prompt unchanged. Invalid outputs remain in the denominator, and plan feasibility and support are recomputed from the immutable evaluator ledger. These controls support reproducible comparison on unseen base items, but the taxonomy and oracle are benchmark-defined rather than drawn from an unseen generator or independent deployment study. Record coverage is therefore reported separately from the shared primary metrics.

\section{WirelessOpsAgent Architecture}
\label{sec:design}

WirelessOpsAgent puts the benchmark check inside the agent loop.

\subsection{Closed Loop Architecture}
\begin{figure*}[!t]
\centering
\includegraphics[width=0.90\textwidth]{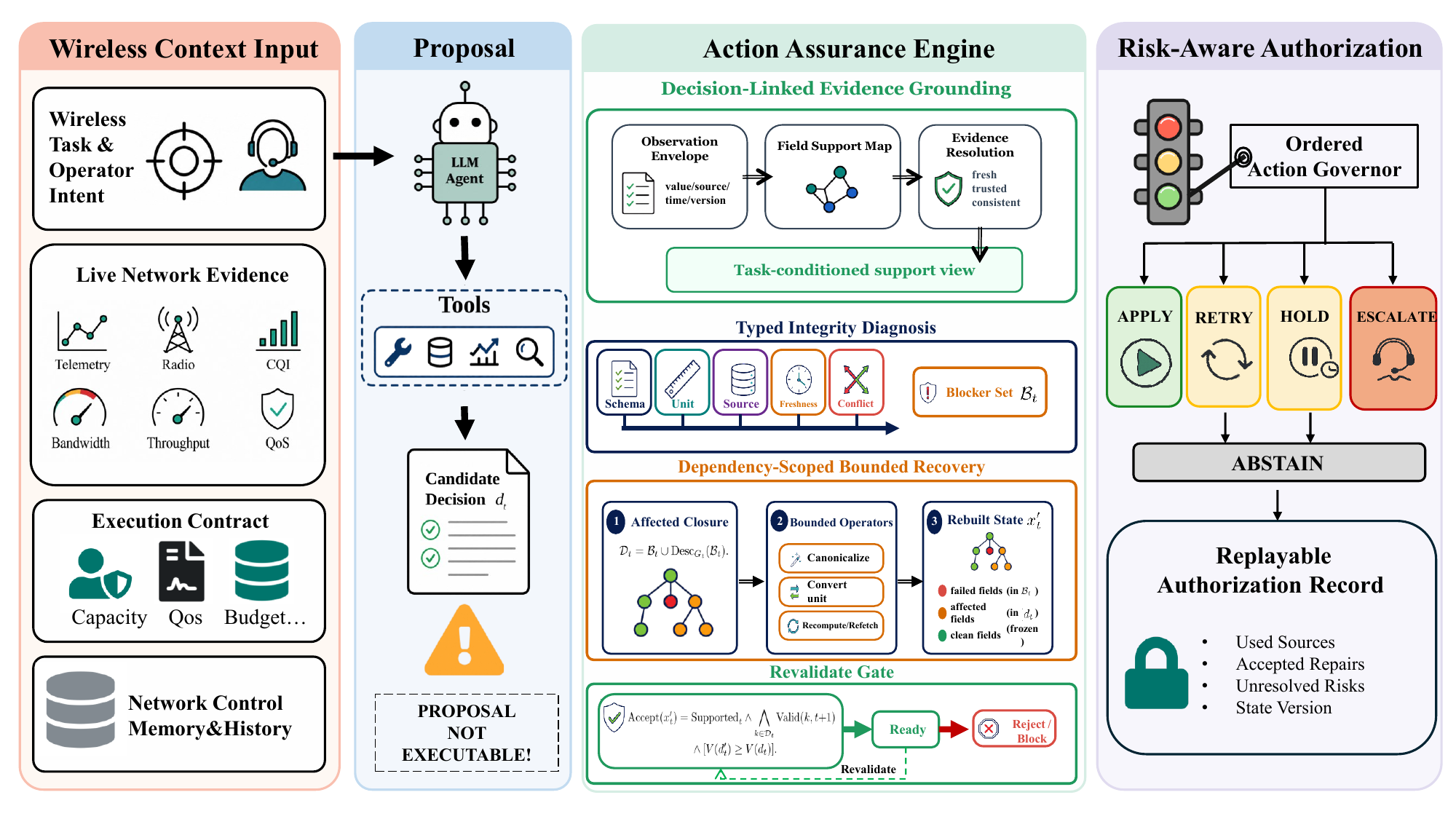}
\caption{WirelessOpsAgent binds evidence, checks a proposed decision, repairs affected fields, and authorizes the resulting action.}
\label{fig:architecture}
\end{figure*}

WirelessOpsAgent is organized around input binding and output authorization. Given a wireless task, a network observation snapshot, and an execution contract, the first stage preserves the exposed observations in an immutable ledger and constructs a lineage preserving canonical view for proposal generation. Input binding secures the observations presented to the model, but generation occurs afterward and can introduce new errors. The model may misstate a unit, omit a required schema field, produce an incomplete allocation or mobility plan, or select APPLY despite unresolved conflict. A nominally correct answer may therefore remain operationally unsafe. WirelessOpsAgent treats the answer, intended action, and plan as a proposal, binds its fields to their dependencies, and checks plan completeness, network constraints, and support. Recoverable dependencies are repaired and revalidated before the action governor releases or blocks execution. This proposal interface can wrap an existing network agent without changing its internal reasoning. Fig.~\ref{fig:architecture} gives the overall flow.

\subsection{Decision Linked Evidence Graph}
Authorization first requires an evidence graph. Each critical field $k$ is represented by its value, schema, unit, source identifier, provenance, event time, ingest time, freshness limit, quality state, and parent identifiers. WirelessOpsAgent maintains three linked views of this information. The raw state preserves observations exactly as exposed by tools. The normalized state maps aliases to canonical task fields and converts compatible units without discarding lineage. The decision state parses $d_t$ into task fields and attaches the dependencies that support them, such as throughput depending on bandwidth and CQI, or QoS depending on throughput and the minimum service rate.

These links form an evidence graph from raw observations to the candidate action. They prevent normalization from hiding a source and prevent a newly computed value from being treated as fresh when one of its parents has expired. Authorization therefore follows dependency edges rather than checking only the final text.

\subsection{Typed Integrity Diagnosis}
Given these links, the integrity monitor evaluates each critical field $k$ at time $t$ using five explicit checks:
\begin{equation}
\begin{aligned}
\mathrm{Valid}(k,t)={}&\mathrm{SchemaOK}(k)\land\mathrm{UnitOK}(k)\\
&\land\mathrm{SourceOK}(k,t)\land\mathrm{Fresh}(k,t)\\
&\land\mathrm{ConflictFree}(k,t).
\end{aligned}
\label{eq:field-validity}
\end{equation}
The five terms correspond to schema validity, unit validity, provenance support, temporal validity, and conflict freedom. Schema and unit checks recognize aliases and compatible conversions such as channel quality index to CQI or Gbps to Mbps. Temporal checks combine event to ingest age, the declared freshness requirement, state version order, and bounded clock skew instead of trusting arrival order alone. Provenance checks require each critical field and each repair to resolve to an existing ledger source. Conflict checks group observations for the same quantity and reject unresolved disagreement or an older event that overwrites a newer trusted value. Domain and cross field checks enforce legal CQI and bandwidth ranges, recompute throughput from authorized inputs, and validate QoS against the resulting rate.

Validation runs at both the field and dependency levels. A successfully canonicalized alias is recorded as a resolved transformation rather than a reason to refuse. A stale parent remains a blocking violation even when its downstream value was computed recently. A tool brownout carries retry eligibility and budget semantics. The monitor therefore produces typed violations rather than a single anomaly score. These types determine both what can be repaired and which blocking action is appropriate when repair fails.

\subsection{Dependency Scoped Bounded Recovery}
To avoid unnecessary blocking or broad rewriting, let $\mathcal{B}_t$ denote the critical fields that fail integrity diagnosis. WirelessOpsAgent limits recovery to these fields and their descendants in the evidence graph:
\begin{equation}
\mathcal{D}_t=\mathcal{B}_t\cup\mathrm{Desc}_{G_t}(\mathcal{B}_t).
\label{eq:repair-scope}
\end{equation}
The agent computes $r_t=\mathrm{Repair}_{\mathcal{B}_t}(x_t,G_t)$, rebuilds the values in $\mathcal{D}_t$, and copies all fields outside this scope unchanged. A repair is accepted only when its changed fields remain supported, pass revalidation, and do not reduce task utility:
\begin{equation}
\begin{aligned}
\mathrm{Accept}(x'_t)={}&\mathrm{Supported}_t
\land \bigwedge_{k\in\mathcal{D}_t}\mathrm{Valid}(k,t{+}1)\\
&\land [V(d'_t)\geq V(d_t)].
\end{aligned}
\label{eq:repair-acceptance}
\end{equation}
Here, $V$ is the task verifier, and $\mathrm{Supported}_t$ requires every changed field and repair to resolve to recorded evidence. This scoped update permits an aliased field to be renamed, a compatible unit converted, or a derived descendant recomputed while unrelated clean fields remain unchanged. A rejected repair leaves the violation blocked. Replacing newer evidence with stale, conflicting, or memory derived state is counted as an unsafe overwrite.

After an accepted repair, WirelessOpsAgent uses the rebuilt state rather than the original candidate. This loop permits a current secondary source, canonical conversion, verified recomputation, or bounded retry to restore authorization. Unresolved stale, unavailable, or conflicting evidence remains blocked, so answer plausibility cannot override source admissibility.

Memory is treated as an untrusted source rather than a privileged answer cache. It is considered only when current public tools leave a critical field missing or unsafe, and it must pass the same task, freshness, provenance, and dependency checks as a tool observation. Current valid evidence dominates memory. The design therefore separates retrieval utility from admission authority and permits an ablation with memory disabled.

\subsection{Risk Aware Authorization and Replayable Records}
After validation and bounded recovery, authorization decides whether the proposal can be released. A candidate is ready only when its evidence, network constraints, and authorization record all pass. Formally, $\mathrm{Ready}_t=\mathrm{FieldsValid}_t\land\mathrm{ConstraintsPass}_t\land\mathrm{RecordComplete}_t$, where $\mathrm{FieldsValid}_t$ means that every critical field satisfies Eq.~\eqref{eq:field-validity}. The action governor then evaluates the following ordered policy from top to bottom:
\begin{equation}
a_t=
\begin{cases}
\mathrm{APPLY}, & \mathrm{Ready}_t=1,\\
\mathrm{ESCALATE}, & \text{ambiguity or conflict remains},\\
\mathrm{RETRY}, & \text{a required tool can be retried},\\
\mathrm{HOLD}, & \text{required state is not yet current},\\
\mathrm{ABSTAIN}, & \mathrm{otherwise}.
\end{cases}
\label{eq:action-policy}
\end{equation}
APPLY is selected only when every critical field, network constraint, and record consistency check passes. ESCALATE takes precedence when ambiguous schema, unresolved conflict, or independent mixed risks remain. RETRY is used when a required source exposes a retryable protocol state and finite backoff, while HOLD prevents execution until time sensitive evidence becomes current. ABSTAIN is the fallback when no supported action is available.

The authorization record $z_t$ contains the normalized action, state version, structured decision, field level used sources, blocked sources, integrity results, repair actions, unresolved risks, and provenance coverage. Hard integrity failures take precedence over verifier confidence. The record preserves both the violations before repair and the support graph after repair, allowing the evaluator to replay why a candidate was released or blocked. A record is faithful only when its claims agree with the public observation ledger.

\section{Experimental Results}
\label{sec:evaluation}
\begin{table*}[t]
	\centering
	\caption{Metric definitions. $N$ includes invalid outputs. $N_A$ counts reference APPLY episodes. $\mathcal{A}_{\mathrm{ref}}$ contains the four reference actions. $S_i$ marks a feasible APPLY without violations. $U_i$ marks an APPLY with a hard blocker. Readiness Score (OPS) averages $d_i,u_i,s_i,e_i$ (decision, utility, state, record), with $g_i=0$ on hard violations. Blockers cover evidence, feasibility, authorization, and unsafe overwrite.}
	\label{tab:metric-glossary}
	\begingroup
	\setlength{\tabcolsep}{3pt}
	\renewcommand{\arraystretch}{1.12}
	\scriptsize
	\begin{tabular}{@{}p{0.18\textwidth}p{0.32\textwidth}p{0.44\textwidth}@{}}
		\toprule
		Metric & Formula & Plain language meaning \\
		\midrule
		Task Score & $\frac{1}{N}\sum_i \mathrm{Score}_i$ & Measures nominal wireless task quality without considering whether the action is safe to execute. \\
		Readiness Score (OPS) & $\frac{1}{N}\sum_i g_i(d_i+u_i+s_i+e_i)/4$ & Summarizes overall episode quality after hard violations are gated. It is interpreted together with the primary metrics. \\
		Exact Action Accuracy & $\frac{1}{N}\sum_i\mathbf{1}[\hat a_i=r_i]$ & Measures the fraction of episodes whose predicted action exactly matches the reference action. \\
		Action Choice Macro F1 & $\frac{1}{4}\sum_{c\in\mathcal{A}_{\mathrm{ref}}}\mathrm{F1}_c$ & Provides a balanced action diagnostic for the matched ablation replay. \\
		Safe APPLY Recall & $\frac{1}{N_A}\sum_{r_i=\mathrm{APPLY}}S_i$ & Measures the fraction of reference APPLY episodes that produce a complete, feasible, and violation free APPLY decision. \\
		Unsafe APPLY Rate & $\frac{1}{N}\sum_i U_i$ & Measures the fraction of all episodes in which the agent selects APPLY despite a hard blocker. \\
		Safe APPLY Utility & $\frac{1}{N_A}\sum_{r_i=\mathrm{APPLY}}S_i\mathrm{Score}_i$ & Measures the WirelessBench task utility retained by complete, feasible, and violation free APPLY decisions. \\
		\bottomrule
	\end{tabular}
	\endgroup
\end{table*}
The evaluation moves from aggregate outcomes to condition and ablation analyses, keeping each result tied to the claim it supports.

\subsection{Experimental Protocol}
We compare WirelessOpsAgent with three baselines: Direct, WirelessAgent++ + contract, and CRITIC style. Direct is the single pass comparator on the common label free execution state interface. WirelessAgent++ + contract augments the official source and paper workflow with the same explicit action contract~\cite{tong2026wirelessagentautomatedagenticworkflow}, while CRITIC style adds tool mediated checking~\cite{gou2024critic}. Every method is evaluated on the frozen 600 episode split from Section~\ref{sec:benchmark}. We keep the separately exported Direct minimal/no contract sensitivity outside the headline comparison and report the seven GPT-5.4-mini ablations only in the matched ablation study.

The backbone comparison covers GPT-5.4-mini, Qwen3-8B, and Claude Sonnet 4.6, with 600 episodes for each method on each backbone under one evaluator. Qwen and Claude use the corrected endpoint mapping. The release retains the original recorded labels and source hashes, while each backbone shares a frozen episode inventory across methods. All selected rows are valid and have no runtime errors. The condition and ablation studies keep the GPT-5.4-mini inference configuration fixed and use the same frozen hard violation scorer.

\subsection{Metric Definitions}
\label{sec:benchmark-metrics}
Because outputs contain both task answers and actions, the metrics separate task quality from execution readiness. Our primary operational metrics are Exact Action Accuracy, Safe APPLY Recall, and Unsafe APPLY Rate. Together they separate correct routing, retained executable actions, and authorization failures. Task Score captures nominal answer quality, while Safe APPLY Utility measures how much task utility remains after execution requirements are enforced. OPS is a gated secondary summary, and Action Choice Macro F1 is used only for the matched ablation replay. Table~\ref{tab:metric-glossary} gives the definitions. Invalid responses and model call failures remain in every relevant denominator.

Binary integrity audits are indicator means on their declared split. Confidence intervals are the 2.5th and 97.5th percentiles of 10,000 paired bootstrap mean differences stratified by task. Each stress domain contains 120 episodes, so the equally weighted domain average equals the overall mean.

\subsection{Action Correctness, Safety, and Progress}

\begin{table}[t]
\centering
\caption{Stress results across backbones (600 episodes for each method on each backbone). Safe-R is Safe APPLY Recall over 200 reference APPLY episodes. Unsafe is Unsafe APPLY Rate over all 600 episodes.}
\label{tab:main-results}
\begingroup
\setlength{\tabcolsep}{2.3pt}
\scriptsize
\begin{tabular}{@{}l r r r r r@{}}
\toprule
Method & Task $\uparrow$ & OPS $\uparrow$ & Exact $\uparrow$ & Safe-R $\uparrow$ & Unsafe $\downarrow$ \\
\midrule
\multicolumn{6}{c}{\emph{GPT-5.4-mini} ($N=600$)} \\
\midrule
Direct & 0.539 & 0.355 & 0.538 & 25.5\% & 49.0\% \\
WirelessAgent++ + contract & 0.424 & 0.351 & 0.517 & 14.5\% & 49.0\% \\
CRITIC style & 0.324 & 0.555 & 0.560 & 18.5\% & 18.7\% \\
\rowcolor{opsrowgray}
WirelessOpsAgent & \textbf{0.802} & \textbf{0.912} & \textbf{0.972} & \textbf{72.5\%} & \textbf{7.8\%} \\
\midrule
\multicolumn{6}{c}{\emph{Qwen3-8B} ($N=600$)} \\
\midrule
Direct & 0.315 & 0.386 & 0.675 & 22.0\% & 47.5\% \\
WirelessAgent++ + contract & 0.300 & 0.452 & 0.707 & 17.0\% & 36.3\% \\
CRITIC style & 0.236 & 0.489 & 0.632 & 19.0\% & 30.0\% \\
\rowcolor{opsrowgray}
WirelessOpsAgent & \textbf{0.389} & \textbf{0.815} & \textbf{0.983} & \textbf{48.0\%} & \textbf{17.2\%} \\
\midrule
\multicolumn{6}{c}{\emph{Claude Sonnet 4.6} ($N=600$)} \\
\midrule
Direct & 0.508 & 0.076 & 0.370 & 17.0\% & 88.7\% \\
WirelessAgent++ + contract & 0.468 & 0.122 & 0.407 & 12.0\% & 82.2\% \\
CRITIC style & 0.483 & 0.076 & 0.363 & 16.5\% & 88.5\% \\
\rowcolor{opsrowgray}
WirelessOpsAgent & \textbf{0.520} & \textbf{0.878} & \textbf{0.975} & \textbf{66.0\%} & \textbf{10.3\%} \\
\bottomrule
\end{tabular}
\endgroup
\end{table}

Table~\ref{tab:main-results} tests whether proposal checks reduce unsafe execution without collapsing safe progress. WirelessOpsAgent attains Exact Action Accuracy between 0.972 and 0.983, whereas the strongest baseline reaches at most 0.707. Its Unsafe APPLY Rate falls to 7.8\%, 17.2\%, and 10.3\% on GPT-5.4-mini, Qwen3-8B, and Claude Sonnet 4.6. Relative to the safest baseline on each backbone, these values are lower by 10.9, 12.8, and 71.9 percentage points. The safety gain does not eliminate safe progress. Safe APPLY Recall rises to 72.5\%, 48.0\%, and 66.0\%, compared with best baseline values of 25.5\%, 22.0\%, and 17.0\%. WirelessOpsAgent also records the highest Task Score and OPS in every backbone block.

The backbone results also separate answer quality from execution readiness. On Claude Sonnet 4.6, Direct and CRITIC style retain Task Scores near 0.5 but collapse in OPS because they frequently apply under blockers. On Qwen3-8B, WirelessOpsAgent has a lower Task Score than on GPT-5.4-mini but the highest Exact Action Accuracy. The claim is therefore not that the backbone solves every wireless calculation better, but that the assurance loop converts imperfect proposals into safer operational decisions.

A trivial safe policy would lower unsafe APPLY by refusing most executions. The results show a different pattern. WirelessOpsAgent keeps the largest Safe APPLY Recall on every backbone and improves exact routing at the same time, so the safety improvement is coupled with useful progress. The gap between Task Score and OPS in the baselines also explains why nominal answer quality is insufficient: a model can compute a plausible wireless answer and still release it under stale, missing, or conflicting support.

\begin{figure}[!t]
\centering
\includegraphics[width=0.70\columnwidth]{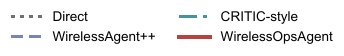}\par
\vspace{-0.4ex}
\subfloat[GPT-5.4-mini]{%
\includegraphics[width=0.84\columnwidth]{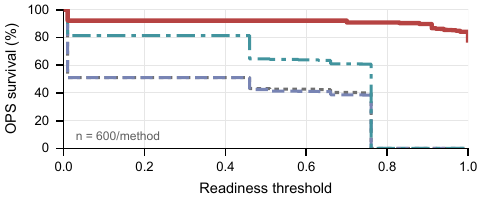}%
}\par
\vspace{-0.8ex}
\subfloat[Qwen3-8B]{%
\includegraphics[width=0.84\columnwidth]{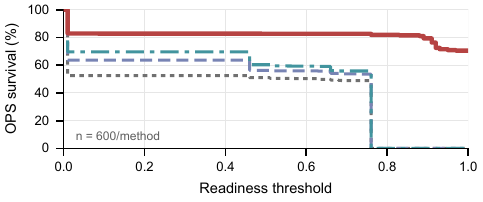}%
}\par
\vspace{-0.8ex}
\subfloat[Claude Sonnet 4.6]{%
\includegraphics[width=0.84\columnwidth]{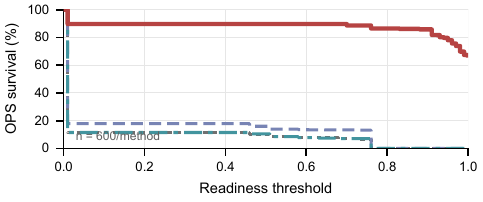}%
}
\caption{OPS survival over 600 episodes for each method on each backbone. The ordinate is the fraction with OPS at least $\tau$.}
\label{fig:ops-survival}
\end{figure}

The OPS survival curves in Fig.~\ref{fig:ops-survival} show that this advantage extends across the episode distribution. At $\tau=0.8$, WirelessOpsAgent retains 90.8\%, 81.8\%, and 86.5\% of episodes for the three backbones, while every baseline curve has reached zero. OPS is supporting evidence because it includes record faithfulness. The primary conclusion remains grounded in exact actions, safe progress, and unsafe execution.

\begin{table}[!t]
\centering
\caption{APPLY decisions before and after repair. APPLY is positive. True positives (TP), false positives (FP), false negatives (FN), true negatives (TN), false positive rate (FPR), and accuracy (Acc.) are computed over the 600 episode split.}
\label{tab:apply-confusion}
\begingroup
\setlength{\tabcolsep}{1pt}
\scriptsize
\begin{tabular}{@{}llrrrrrrr@{}}
\toprule
Backbone & Stage & TP & FP & FN & TN & Recall & FPR & Acc. \\
\midrule
\multirow{2}{*}{GPT-5.4-mini} & Before repair & 187 & 369 & 13 & 31 & 93.5\% & 92.2\% & 36.3\% \\
& After repair & 192 & 0 & 8 & 400 & 96.0\% & 0.0\% & 98.7\% \\
\midrule
\multirow{2}{*}{Qwen3-8B} & Before repair & 198 & 342 & 2 & 58 & 99.0\% & 85.5\% & 42.7\% \\
& After repair & 199 & 0 & 1 & 400 & 99.5\% & 0.0\% & 99.8\% \\
\midrule
\multirow{2}{*}{Claude Sonnet 4.6} & Before repair & 194 & 357 & 6 & 43 & 97.0\% & 89.2\% & 39.5\% \\
& After repair & 194 & 0 & 6 & 400 & 97.0\% & 0.0\% & 99.0\% \\
\bottomrule
\end{tabular}
\endgroup
\end{table}

One possible explanation for the lower Unsafe APPLY Rate is indiscriminate blocking. Table~\ref{tab:apply-confusion} rules out that explanation at the binary authorization level. After repair, false APPLY decisions fall to zero on all 400 blocking reference episodes for each backbone, while APPLY recall remains between 96.0\% and 99.5\%. This audit is deliberately narrower than Unsafe APPLY Rate, which also counts hard violations when APPLY is the reference action.

The before and after rows expose the source of this gain. Before repair, all backbones propose APPLY for most blocking episodes, with FPR between 85.5\% and 92.2\%. After diagnosis and scoped recovery, FPR drops to zero while recall remains high. The remaining false negatives are therefore not evidence of generic rejection. They are cases where the rebuilt proposal still lacks a complete, supported, and feasible APPLY decision.

\begin{figure}[!t]
\centering
\includegraphics[width=0.78\columnwidth]{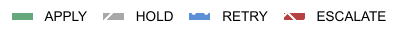}\par
\vspace{-0.3ex}
\subfloat[GPT-5.4-mini]{%
\includegraphics[width=0.86\columnwidth]{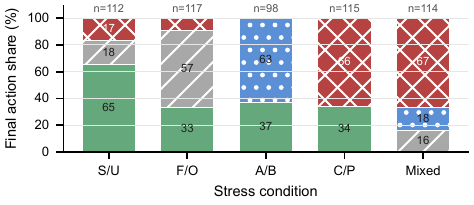}%
}\par
\vspace{-0.8ex}
\subfloat[Qwen3-8B]{%
\includegraphics[width=0.86\columnwidth]{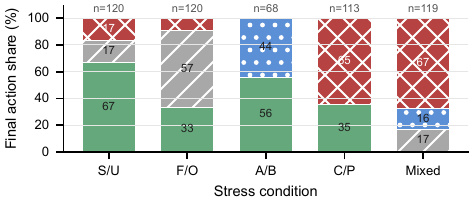}%
}\par
\vspace{-0.8ex}
\subfloat[Claude Sonnet 4.6]{%
\includegraphics[width=0.86\columnwidth]{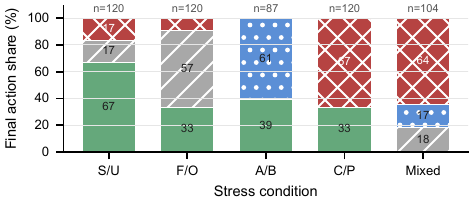}%
}
\caption{Final routing of proposals initially marked APPLY. Bars are normalized by the count shown above them. The conditions are schema/unit (S/U), freshness/order (F/O), availability/brownout (A/B), and conflict/provenance (C/P).}
\label{fig:apply-routing}
\end{figure}

We next trace where blocked APPLY proposals go. Fig.~\ref{fig:apply-routing} shows condition specific routing rather than a generic rejection response. Across backbones, freshness/order faults send 57\% of initial APPLY proposals to HOLD, availability/brownout faults send 44--63\% to RETRY, and conflict/provenance faults send 66--67\% to ESCALATE. Mixed evidence invokes all three alternatives. The governor therefore maps distinct failures to distinct operational responses.

This routing matters because the blocking actions have different operational meanings. HOLD waits for a current state, RETRY reissues a bounded telemetry request, and ESCALATE asks for human resolution when ambiguity or conflict remains. Exact action scoring therefore tests more than whether an agent refuses APPLY. It tests whether the refusal is useful for the operator and consistent with the evidence fault.

\begin{figure*}[!t]
\centering
\includegraphics[width=\textwidth]{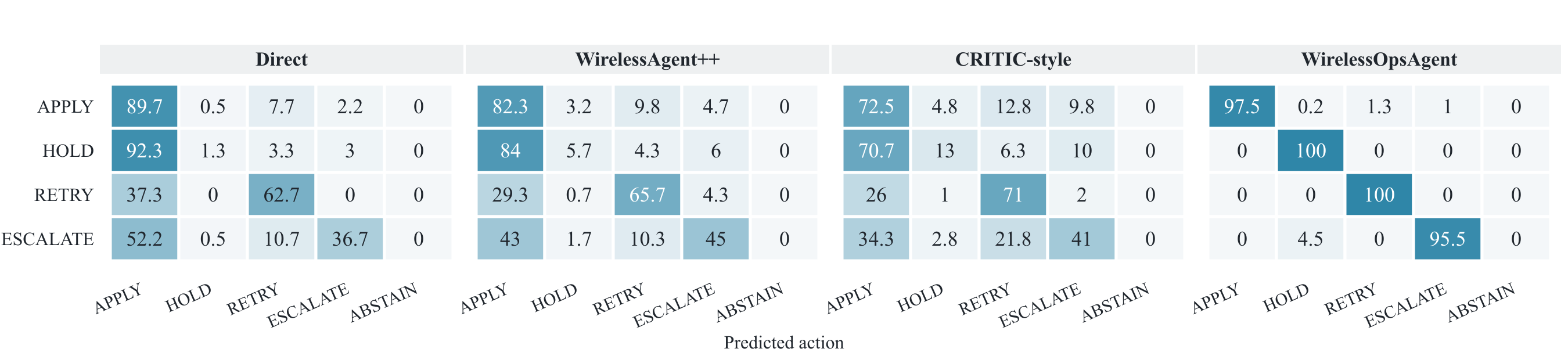}
\caption{Row normalized action confusion matrices averaged over three 600 episode backbone runs. Rows are reference actions. Columns include the four benchmark actions and fallback ABSTAIN.}
\label{fig:action-confusion}
\end{figure*}

Fig.~\ref{fig:action-confusion} then places these repairs in the full four action space. All three baselines over select APPLY on blocking cases, despite differences in their remaining errors. WirelessOpsAgent is nearly diagonal across backbones. Its limited residual mass shifts mainly from APPLY toward RETRY or ESCALATE, or from ESCALATE toward HOLD, which is consistent with conservative handling rather than unguarded execution.

\subsection{Condition Diagnostics}

\begin{table}[t]
\centering
\caption{WirelessOpsAgent results by condition, macro averaged over three task families (120 episodes each). Labels follow the row level execution contract. Dashes indicate no reference APPLY episodes.}
\label{tab:scenario}
\begingroup
\setlength{\tabcolsep}{2.2pt}
\scriptsize
\begin{tabular}{@{}l r r r r r@{}}
\toprule
Condition & $n$ & Exact $\uparrow$ & Safe util. $\uparrow$ & Safe rec. $\uparrow$ & Unsafe $\downarrow$ \\
\midrule
Schema/unit & 120 & 0.967 & 0.752 & 0.850 & 0.067 \\
Freshness/order & 120 & 0.917 & 0.917 & 0.975 & 0.000 \\
Availability/brownout & 120 & 0.983 & 0.776 & 0.950 & 0.000 \\
Conflict/provenance & 120 & 0.992 & 0.000 & 0.000 & 0.325 \\
Mixed evidence & 120 & 1.000 & - & - & 0.000 \\
\bottomrule
\end{tabular}
\endgroup
\end{table}

Having established the aggregate pattern, we examine where it holds and where the evaluator needs care. Table~\ref{tab:scenario} reports zero frozen unsafe APPLY decisions for freshness/order, availability/brownout, and mixed evidence. Schema/unit stress remains harder, with an Unsafe APPLY Rate of 0.067, but still retains 0.752 Safe APPLY Utility.

The conflict/provenance value of 0.325 is confounded by the frozen evaluator and should not be read as a material conflict failure rate. A post freeze audit found that nonmaterial hard negatives were scored as conflicts, which marked 39 of 40 correct WirelessOpsAgent APPLY decisions and 22 of 40 CRITIC style decisions unsafe. Neither method selected APPLY on the 80 material conflict positives. We retain the frozen score for reproducibility and exclude this slice from claims about material conflict handling. Across the other four domains, WirelessOpsAgent produces 8 unsafe APPLY decisions among 480 episodes, compared with 89 for CRITIC style.

\begin{figure}[!t]
\centering
\subfloat[GPT-5.4-mini]{%
\includegraphics[width=0.88\columnwidth]{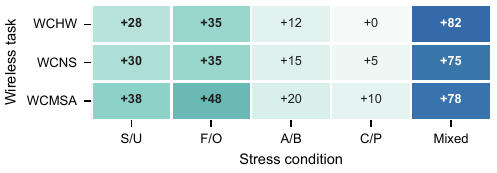}%
}\par
\vspace{-0.8ex}
\subfloat[Qwen3-8B]{%
\includegraphics[width=0.88\columnwidth]{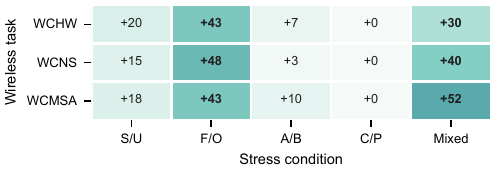}%
}\par
\vspace{-0.8ex}
\subfloat[Claude Sonnet 4.6]{%
\includegraphics[width=0.88\columnwidth]{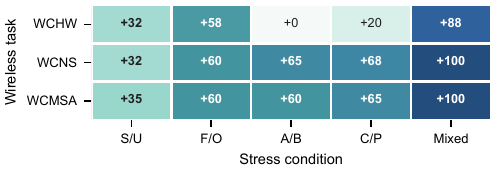}%
}\par
\vspace{-0.2ex}
\includegraphics[width=0.76\columnwidth]{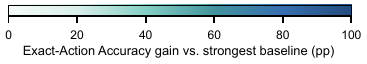}
\caption{Exact Action Accuracy gain over the strongest baseline in each backbone, task, and condition cell (percentage points, 40 episodes per method). The conditions are schema/unit (S/U), freshness/order (F/O), availability/brownout (A/B), and conflict/provenance (C/P).}
\label{fig:task-condition-gain}
\end{figure}

The cell view in Fig.~\ref{fig:task-condition-gain} checks whether the aggregate gain depends on a small subset of tasks or backbones. WirelessOpsAgent exceeds the strongest baseline in 40 of 45 cells defined by backbone, task, and condition, and ties in the other five. No cell is negative. Gains are largest under mixed evidence, where a correct decision must reconcile several simultaneous blockers, while the ties occur where a baseline already selects every reference action correctly.

The cell pattern strengthens the benchmark claim. Gains appear in WCHW, WCNS, and WCMSA rather than only in calculation heavy WCHW cases, so the effect is not confined to one task family. The zero gain cells are ceiling cases in which a baseline already matches the reference action. This makes the aggregate improvement less vulnerable to a single backbone, task, or stress condition.

\subsection{Ablations and Faithfulness}

\begin{table}[t]
\centering
\caption{Matched ablation on 600 GPT-5.4-mini episodes. Macro F1 averages the four reference actions. Full reproduces the main result.}
\label{tab:ablation}
\begingroup
\setlength{\tabcolsep}{2.4pt}
\scriptsize
\begin{tabular}{@{}l r r r r@{}}
\toprule
Variant & Macro F1 $\uparrow$ & Safe util. $\uparrow$ & Safe rec. $\uparrow$ & Unsafe $\downarrow$ \\
\midrule
WirelessOpsAgent full & 0.972 & \textbf{0.639} & \textbf{0.725} & \textbf{0.078} \\
Monitor only & \textbf{0.983} & 0.224 & 0.250 & 0.250 \\
No integrity monitor & 0.216 & 0.632 & 0.715 & 0.688 \\
No action governor & 0.219 & \textbf{0.639} & \textbf{0.725} & 0.693 \\
No utility repair & 0.964 & 0.224 & 0.245 & 0.230 \\
No schema and unit normalizer & 0.889 & \textbf{0.639} & \textbf{0.725} & 0.157 \\
No conflict and freshness checks & 0.514 & \textbf{0.639} & \textbf{0.725} & 0.355 \\
\bottomrule
\end{tabular}
\endgroup
\end{table}

Table~\ref{tab:ablation} explains which parts of the loop produce both safety and progress. The integrity monitor and action governor form the main authorization path. Removing either reduces Action Choice Macro F1 from 0.972 to about 0.22 and raises Unsafe APPLY Rate from 0.078 to about 0.69. Conflict and freshness checks provide a second safety layer, as their removal raises unsafe execution to 0.355 and lowers Macro F1 to 0.514.

Safety checks alone are not enough. The monitor only variant attains 0.983 Macro F1, yet its Safe APPLY Recall is only 0.250 and its Unsafe APPLY Rate is 0.250. Utility repair recovers safe progress from recoverable inputs. Without it, Safe APPLY Utility falls from 0.639 to 0.224 and Safe APPLY Recall from 0.725 to 0.245. Schema and unit normalization addresses a narrower but measurable risk, with its removal doubling unsafe execution from 0.078 to 0.157. All variants preserve clean fields at 1.000 and record zero unsafe overwrites, so the differences arise from authorization and recovery rather than broad state rewriting.

The ablations therefore separate three failure modes: missing monitors hide blockers, missing governors release unsupported actions, and missing repair sacrifices recoverable safe progress.

Taken together, the ablations rule out three simpler explanations: a conservative reject policy, a stronger prompt, or a formatting artifact. The full system is the only variant that combines low Unsafe APPLY Rate with high Safe APPLY Utility. The zero unsafe overwrite rate further shows that recovery is scoped because it repairs compromised dependencies without rewriting fields that already satisfy the contract.

\section{Limitations}
WirelessOptBench is a controlled benchmark rather than a live network deployment study. Its stress domains are designed to test execution state reasoning, so they should not be interpreted as production fault frequencies or an exhaustive catalog of wireless operations. The backbone comparison covers three LLMs, and the ablations use GPT-5.4-mini. WirelessOpsAgent also assumes that exposed telemetry metadata and provenance fields are correctly reported. Actuator security and concurrent control are outside the present evaluation.

\section{Conclusion}
Reliable wireless network operations require agents to determine not only what action to take, but also whether that action is supported by the current network state. This paper introduced WirelessOptBench to evaluate this requirement and WirelessOpsAgent to establish action assurance before execution. The results show that evidence grounding, bounded recovery, and revalidation can improve operational decision making under imperfect network observations. Together, the benchmark and agent provide a foundation for developing wireless network agents whose actions are effective, dependable, and auditable.

\clearpage
\bibliographystyle{IEEEtran}
\bibliography{refs}

\end{document}